\documentstyle{article}
\begin{document}
\LARGE
\begin{center}
\bf  Pair Creation of Black Holes in Anti-de Sitter Space
Background (I)
\vspace*{0.6in}
\normalsize \large \rm

Wu Zhong Chao

Dept. of Physics

Beijing Normal University

Beijing 100875, China

(Oct. 4, 1998)

\vspace*{0.4in}
\large
\bf
Abstract
\end{center}
\vspace*{.1in}
\rm
\normalsize
\vspace*{0.1in}

For a spherically symmetric vacuum model with a negative
cosmological constant,  a
complex constrained instanton is considered as the seed for
the quantum pair creation of 
Schwarzschild-anti-de Sitter black holes. The relative creation
probability
is found to be the exponential of the negative
of  the black hole
entropy. The black hole entropy is known to be one quarter of the
black hole horizon area. In the absence of a general no-boundary
proposal for
open creation, 
the constrained instanton approach is used in treating both the
open
and closed pair creations of black holes.

\vspace*{0.3in}

PACS number(s): 98.80.Hw, 98.80.Bp, 04.60.Kz, 04.70.Dy


\vspace*{0.3in}

e-mail: wu@axp3g9.icra.it

\pagebreak

\vspace*{0.3in}

In the No-Boundary Universe, the wave function of a closed
universe is  defined as
a path integral over all compact 4-metrics with matter fields
[1]. The dominant contribution to
the path integral is from the stationary action solution. At
the $WKB$ level, the wave function
can be written as
\begin{equation}
\Psi \approx e^{- I},
\end{equation}
where $I= I_r + iI_i$  is the complex action of the solution.

The imaginary part $I_i$ and real part $I_r$ of the
action represent the Lorentzian
and Euclidean evolutions in real time and imaginary time,
respectively. When their orbits are intertwined, they are
mutually perpendicular in the configuration space with the
supermetric. The probability of a
Lorentzian orbit remains constant during the evolution. One can
identify the probability, not only
as the probability of the universe created, but also as
the probabilities for other Lorentzian
universes obtained through an analytic continuation from it [2].

An instanton is defined as a stationary action orbit and
satisfies the Einstein equation everywhere, and it is
the seed for the creation of the universe. However, very few
regular instantons exist. The
framework of the No-Boundary Universe is much wider than that of
the instanton theory. Therefore, in order not to exclude many
interesting phenomena from
the study, one has to appeal
to the concept of constrained instantons [3]. Constrained
instantons are the orbits
with an action which is stationary
under some restriction. The restriction can be imposed
on a spacelike 3-surface of the created Lorentzian
universe. This restriction is that the 3-metric and matter
content are given at the 3-surface. The relative creation
probability from the
instanton is the exponential of the negative of the
real part of the instanton action.

The usual prescription for finding a constrained instanton is to
obtain a complex solution to the Einstein equation and other
field equations in the complex domain of spacetime coordinates.
If there is no singularity in a compact sector of the solution,
then the sector is considered as an instanton. If there exist
singularities in the sector, then the action of the sector is not
stationary. The action may only be stationary with respect to the
variations under some restrictions mentioned above. If this is
the case, then the sector is a constrained gravitational
instanton. To find the constrained instanton, one has to closely
investigate the singularities. The stationary action condition is
crucial to the validation of the $WKB$ approximation. We are
going to work at the $WKB$ level for
the problem of quantum creation of a black hole pair.

A main unresolved problem in quantum cosmology is to
generalize the no-boundary proposal for an open universe. While a
general prescription  is not available, one can still use 
analytic continuation to obtain the $WKB$ approximation to the
wave function for open universes with some kind of symmetry.

The most symmetric space is the $S^4$ space with $O(5)$ symmetry,
or the four-sphere,
\begin{equation}
ds^2 = d\tau^2 + \frac{3}{\Lambda} \cos^2 \left (
\sqrt{\frac{\Lambda}{3}} \tau \right ) (d \chi^2
+ \sin^2 \chi (d \theta^2 + \sin^2 \theta d \phi^2)) ,
\end{equation}
where $\Lambda$ is a positive cosmological constant.  One can
obtain the de Sitter space or anti-de
Sitter space by the substitution $\tau = it$ or $ \chi = i\rho$,
respectively, The signature of the de Sitter
space is $(-, +, +, +)$ and that for the anti-de Sitter space is
$(+,
-, -, -)$. This signature associated
with the anti-de Sitter space is reasonable, since the relative
sign of the cosmological constant is
implicitly changed by the analytic continuation. If one prefers
the usual signature of the anti-de
Sitter space, then he could start from the four-sphere with the
signature of $(-, -, -, -)$, instead of (2) [4].

One can reduce the symmetry to make the model more realistic.
This is the $FLRW$ space with $O(4)$ symmetry,
\begin{equation}
ds^2 = d\tau^2 + a^2(\tau )(d \chi^2 + \sin^2 \chi (d \theta^2 +
\sin^2 \theta d \phi^2)),
\end{equation}
where $a(\tau )$ is the length scale of the homogeneous
3-surfaces and $a(0) = 0$. One can apply combined analytic
continuation [2]
\begin{equation}
\chi = i\rho,
\end{equation}
\begin{equation}
t = -i\tau.
\end{equation}
This results in an open $FLRW$ universe. The study of the
perturbation modes around this background,
strictly following the no-boundary philosophy, is waiting for a
general proposal for
the quantum state of an open universe. If one includes 
``realistic'' matter fields in the model, then the instanton is
not regular. However, the singular instanton can be interpreted
as a constrained instanton [5].

In this paper, we try to reduce the symmetry further, that is to
investigate models with $O(3)$, or
spherical symmetry. We shall consider the vacuum model with a
cosmological constant first. We
shall show that the constrained instanton will lead to a pair
creation of black holes
in de Sitter or anti-de Sitter space
background. The  solution to the Einstein equation is written
\begin{equation}
ds^2 = \Delta d\tau^2 + \Delta^{-1} dr^2 + r^2( d\theta^2 +
\sin^2 \theta d\phi^2 ),
\end{equation}
\begin{equation}
\Delta = 1 -\frac{2m}{r} -  \frac{\Lambda r^2}{3},
\end{equation}
where $m$ is an integral constant.  One can make a factorization 
\begin{equation}
\Delta = - \frac{\Lambda}{3r}
(r - r_0)(r - r_2)(r - r_3).
\end{equation}

The  surface gravity $\kappa_i$ of $r_i$ is [6] 
\begin{equation}
\kappa_i = \frac{\Lambda}{6r_i}\prod_{j = 0,2,3, \;\; (j
\neq i)}(r_i - r_j).
\end{equation}

If $\Lambda$ is positive and $0 \le m \le m_c =\Lambda^{-
1/2}/3$,
then $r_2$ and $r_3$ are
real. There exist regular instantons $S^4$ and $S^2 \times S^2$
for the cases $m = 0$ and $m = m_c$, respectively. The case $m =
0$ leads to the creation of a universe without a black hole and
the case $m = m_c$ leads to the creation of a universe with a
pair of maximal black holes [7]. For the general case, one can
make a
constrained instanton as follows. The constrained
instanton is the seed for the quantum creation of a
Schwarzschild-de
Sitter black hole pair, or a sub-maximal black hole pair [3], and
$r_2$
and $r_3$ become the black hole and cosmological horizons for the
holes created. 

One can have two cuts at $\tau = \pm \Delta \tau /2$ between the
two horizons. Then the $f_2$-fold cover around the black hole
horizon $r = r_2$ turns the $(\tau - r)$ plane into a cone with a
deficit angle $2\pi (1 - f_2)$ there. In a similar way one
can have an $f_3$-fold cover at the cosmological horizon. In
order to form a fairly symmetric Euclidean manifold, one can glue
these two cuts under the condition 
\begin{equation}
f_2 \beta_2 + f_3 \beta_3 = 0,
\end{equation}
where $\beta_2 = 2\pi \kappa_2^{-1}$ and $\beta_3 = 2\pi
\kappa_3^{-1}$ are the periods of $\tau$ that avoid conical
singularities in compacting the Euclidean spacetime
at these two horizons, respectively. The absolute values of their
reciprocals are the
Hawking temperature and the Gibbons-Hawking temperature.
If $f_2$ or $f_3$ is different from $1$ (at least one should
be,
since the two periods are different for the sub-maximal black
holes), then the cone at the black hole or cosmological horizon
will have an extra contribution to the action of the instanton.
After the transition to Lorentzian spacetime, the conical
singularities will only affect the real part of the phase of the
wave function, i.e. the probability of the black hole creation.

The action of the gravitational field is
\begin{equation}
I = - \frac{1}{16 \pi} \int_M (R - 2 \Lambda ) - \frac{1}{8 \pi}
\oint_{\partial M } K,
\end{equation}
where $R$ is the scalar curvature of the spacetime $M$, and $K$
is the trace of the second form of the boundary $\partial M$.

The extra contribution due to the conical singularities can be
considered as the degenerate form of the surface term in the
action (11) and can be written as follows:
\begin{equation}
I_{i,deficit} = - \frac{1}{8 \pi}\cdot 4\pi r_i^2\cdot 2\pi
(1 - f_i).\;\;\; (i = 2, 3)
\end{equation}

The volume term of the action for the instanton can be calculated
\begin{equation}
I_{vol} = -\frac{\Lambda}{6} (r^3_3 - r^3_2) f_2 \beta_2.
\end{equation}

Using eqs. (10) - (13), one obtains the total action
\begin{equation}
I = - \pi (r^2_2 + r^2_3).
\end{equation}
This is one quarter of the negative of the sum of the two horizon
areas. One quarter of the sum is the total entropy of the
universe. 
 
It is remarkable to note that the action is independent of the
choice of $f_2$ or $f_3$. Our manifold satisfies the Einstein
equation everywhere except for the two horizons at the equator.
The equator is two joint sections $\tau = consts. $ passing 
these horizons. It divides the instanton into two halves. The
Lorentzian metric of the black hole pair created can be obtained
through an analytic continuation of the time coordinate from
an imaginary to real value at the equator.  
We can impose the  restriction that the 3-geometry
characterized
by the parameter $m$ is given at the equator, i.e. the transition
surface.
The parameter $f_2$ or $f_3$ is the only degree of freedom left,
since the field equation holds elsewhere. Thus, in order to check
whether we get a stationary action solution for the given
horizons, one only needs to see whether the above action is
stationary with respect to this parameter. Our result (14) shows
that our gravitational action has a stationary action and the
manifold is qualified as a constrained instanton. The exponential
of the negative of the action can be
used for the $WKB$ approximation to the probability.

Eq. (14) also implies that no matter which value of $f_2$ or
$f_3$ is
chosen,
the same black hole should be created with the same probability.
Of course, the most dramatic case  is the creation of a universe
from no volume, i.e. $f_2
=f_3 = 0$. 

From eq. (14) it follows that the relative probability of the
pair
creation of black holes in the de Sitter background is 
the exponential of the total entropy of the universe [3] [8]. 

The independence of the action from the imaginary time period
$\beta$,
$f_2 \beta_2$ for our case,
has some interesting consequences. In gravitational
thermodynamics  
the partition function $Z$ is identified with the path
integral under the constraints. Its $WKB$ approximation
is equivalent to the contribution of the background excluding the
fluctuations. 
At this level one has
\begin{equation}
Z = \exp -I.
\end{equation}
The entropy $S$ can be obtained
\begin{equation}
S = - \frac{\beta \partial}{\partial \beta} \ln Z + \ln Z = - I.
\end{equation}
Thus, the condition that $I$ is independent from $\beta$ implies
that the entropy is the negative of the action. For compact
regular instantons, the fact that the entropy is equal to the
negative of the action is shown using different arguments in [9].

One can study quantum no-boundary states of scalar and spinor
fields in
this model. It turns out that these fields are in thermal
equilibrium with the background. The
associated temperature is the reciprocal of the period as
expected, and it can take an arbitrary value [10].

In quantum gravity the quantum state can be represented by a
matrix density. Apparently, the state associated with our
constrained instanton is  an eigenstate of
the entropy operator, instead of the 
temperature operator, as previously thought.

Now, let us discuss the case of $\Lambda < 0$. One is interested
in the probability of pair creation
of Schwarzschild-anti-de Sitter black holes. The universe is
open.
Hence, our key point is to find a
complex solution which has both the universe as its Lorentzian
sector and a compact sector as the seed for the creation, i.e.
the constrained instanton. The real part of its
action will determine the creation probability.

The metric of the constrained instanton takes the same form as
eq. (6). However, two zeros of $\Delta$ become complex
conjugates.
One can define
\begin{equation}
\gamma \equiv \frac{1}{3} \mbox{arcsinh} (3m
|\Lambda |^{1/2})),
\end{equation}
and then one has
\[
r_2 = 2\sqrt{\frac{1}{|\Lambda|}} \sinh \gamma ,
\]
\begin{equation}
r_3 = \bar{r}_0 = \sqrt{\frac{1}{|\Lambda |}} (  - \sinh \gamma -
i \sqrt{3} \cosh \gamma ).
\end{equation}

One can build a complex constrained instanton using the sector
connecting $r_0$ and $r_3$. Since
$r_0$ and $r_3$ are complex conjugates, the real part of $r$ on
the sector is constant, and the
range of the imaginary part runs between $ \pm
i\sqrt{\frac{3}{|\Lambda |}}  \cosh \gamma $. The surface
gravities $\kappa_0$ and $\kappa_3$ are
complex conjugates too.
Following the procedure of constructing the constrained
gravitational instanton for the case
$\Lambda > 0$, we can use complex folding parameters $f_0$ and
$f_3$ to
cut, fold and glue the
complex manifold with
\begin{equation}
f_0 \beta_0  + f_3 \beta_3 = 0.
\end{equation}

As expected, the action is independent of the parameter $f_0$ or
$f_3$ and
\begin{equation}
I = - \pi ( r_0^2 + r_3^2) = \pi \left (-\frac{6}{\Lambda} +
r^2_2 \right ).
\end{equation}

The action is independent from the choice of the time
identification period.
One can always choose the arbitrary time identification period
to be imaginary, and then the Lorentzian sector in which we are
living is associated with the real time. A special choice of the
imaginary time period will regularize the conical singularity of
the Euclidean sector at the black hole horizon $r_2$. However, we
do not have to do so, since constrained instantons are allowed
in quantum cosmology.

One can obtain the Lorentzian metric from an analytic
continuation
of the time coordinate from an imaginary to real value at the
equator of the
instanton. The equator is two joint $\tau = consts.$ sections
passing through
these horizons. The 3-geometry of the equator can be considered
as the restriction
imposed for the constrained instanton. Again, the independence of
(20) from
the time identification period shows that the manifold is
qualified as a
constrained instanton. 

Therefore, the relative probability of the pair creation of
Schwarzschild-anti-de Sitter black holes, at
the $WKB$ level, is the exponential of the negative of one
quarter of the black hole horizon area,  in contrast to
the case of pair creation of black holes in the de Sitter space
background. One quarter of the black hole horizon area is known
to be the entropy in the Schwarzschild-anti-de Sitter universe
[11].

One may wonder why we choose horizons $r_0$ and $r_3$ to
construct the instanton. One
can also consider those constructions involving $r_2$ as the
instantons. However, the real part
of the action for our choice is always greater
than that of the other choices for the given configuration, and
the wave function or the
probability is determined by the classical orbit
with the greatest real part of the action [1]. When we dealt with
the Schwarzschild-de Sitter case, the choice of the instanton
with $r_2$ and $r_3$ had the greatest action accidentally, but we
did not appreciate this earlier.

Using a standard technique designed for spaces
with spatially noncompact geometries [12] the action of the
Schwarzschild-anti-de Sitter space is evaluated as follows:  The
physical action is defined by the difference between the
action of the space under study and that of a
reference background. The
background can be a static solution to the field equation. From
gravitational thermodynamics, one can
derive the entropy from the action, and it turns out that the
entropy is one quarter of the horizon areas, the same result as
for the closed background [12].

One can also use eq. (16) to derive the ``entropy'' and it is the
negative of the action. For the open creation case,
if one naively interprets the horizon areas as the ``entropy'',
then the ``entropy'' is associated with these two complex
horizons.
Equivalently,  one can say that the action is identical to one
quarter of the black hole horizon area at $r_2$, or the black
hole entropy up to a constant, as we learn in the Schwarzschild
black hole case.

The Hawking temperature is defined as the reciprocal of the
absolute value of the time identification period required to make
the Euclidean manifold regular at the horizon. In the background
subtraction approach for an open universe, if one lifts the
regularity condition at the horizon, or
lets the period take an arbitrary real value, one finds
that the action does depend on the
period and becomes meaningless. However, if we calculate the
action using our complex
constrained instanton, then the action is independent of the
complex period $ \beta$. It is noted that the values of the
action are different for these two methods. For the $O(3)$
symmetric 
case, the beautiful aspect of our approach is that
even in the absence of a general no-boundary proposal for open
universes, we treat the creation of the
closed and the open universes in the same way.

Our treatment of quantum creation of the
Schwarzschild-anti-de Sitter space using the constrained
instanton can
be thought of as a prototype of quantum gravity for an open
system without appealing to the background subtraction approach.

The Schwarzschild black hole case can be thought of as the limit
of our case as we let $\Lambda$ approach  $0$ from below.

The problem of quantum creation of the
Reissner-Nordstr$\rm\ddot{o}$m-anti de
Sitter and the Kerr-Newman-anti de Sitter black holes will be
dealt with in a subsequent paper [13].

\vspace*{0.1in}
\rm

\vspace*{0.1in}

\bf References:

\vspace*{0.1in}
\rm

1. J.B. Hartle and S.W. Hawking, \it Phys. Rev. \rm \bf D\rm
\underline{28}, 2960 (1983).

2. S.W. Hawking and N. Turok, \it Phys. Lett. \rm \bf B\rm
\underline{425}, 25 (1998), hep-th/9802030.

3. Z.C. Wu, \it Int. J. Mod. Phys. \rm \bf D\rm\underline{6}, 199
(1997), gr-qc/9801020.

4. Z.C. Wu,  \it Phys. Rev. \rm \bf D\rm
\underline{31}, 3079 (1985).

5. Z.C. Wu, Beijing preprint, hep-th/9803121.

6. G.W. Gibbons and S.W. Hawking, \it Phys. Rev. \bf D\rm
\underline{15}, 2738 (1977).

7. R. Bousso and S.W. Hawking, \it Phys. Rev. \rm \bf D\rm
\underline{52}, 5659 (1995), gr-qc/9506047.

8. R. Bousso and S.W. Hawking, hep-th/9807148.

9. G.W. Gibbons and S.W. Hawking, \it Commun. Math. Phys. \rm
\underline{66}, 291 (1979).
 
10. Z.C. Wu, Beijing preprint, gr-qc/9712066.

11. S.W. Hawking and D.N. Page, \it Commun. Math. Phys. \rm
\underline{87}, 577 (1983).

12. S.W. Hawking,  in \it General Relativity: An Einstein
Centenary Survey, \rm eds. S.W. Hawking and W. Israel, (Cambridge
University Press, 1979).

13. Z.C. Wu, Beijing preprint, gr-qc/9810012.

\end{document}